# Influence of the Environment on the Effect of Super Resonance in Mesoscale Dielectric Spheres

Igor V. Minin[a], Song Zhou[b], and Oleg V. Minin[a]
[a]Tomsk Polytechnic University, 30 Lenin Ave., Tomsk 634050 Russia
[b]Jiangsu Key Laboratory of Advanced Manufacturing Technology, Faculty of Mechanical and Material Engineering, Huaiyin Institute of Technology, Huai'an 223003, China.

**Abstract**

Dielectric mesoscale spheres have aroused strong interest because of their potential to localize light at deep subwavelength volume and to yield extremal internal magnetic and/or electric field enhancements. Recently, we showed that such particle could support high-order Mie resonance modes with giant field localization and enhancement. Optimizing the internal fields appears as a key challenge for enhancing wave matter interactions in dielectric mesoscale particles. However, a dielectric particle is always located in some medium, and not in a vacuum. Moreover, the question is how much the environment medium affects the internal field intensities enhancement in the super-resonance effect. Based on Mie theory we show for the first time that the presence of the environment leads to a significant decrease in the intensity of the field in the particle. Thus, the study of the effect of super-resonance becomes meaningless without taking into account the environment. However, a greater enhancement of the internal field is found for the blue-shifted Mie size parameter of the sphere when the particle, for example, is in air rather than in vacuum.

**Introduction**

One of the optical effect known as "Photonic Jet" achieved at the shadow surface of the mesoscale dielectric particles (with Mie size parameter $q=kR$, $k$- wavenumber and $R$ – particle radius, $q\sim10$) of arbitrary shape with subwavelength electromagnetic field localization and focusing [1-3]. This effect was extended to the terahertz, plasmonic and acoustic domain [4-6]. Usually, the lateral resolution or beam waist of photonic jet focusing is a better than the simple diffraction limit (i.e. less than half of wavelength) [1,2,7-11].

In 2015 [12] Haong et al. noticed about high-resonance effect using a spherical particle with $q\sim37$ and refractive index $n=1.46$ in application to super-resolution. It was shown that one of the resonant scattering coefficient in the Mie theory was about 20 time higher in magnitude than the other coefficients. This abnormal value of the scattering coefficient described as the constructive interference of the one partial wave inside the wavelength-scaled sphere.

Four years later, in 2019-2020, we reported that lossless dielectric spheres could support so-called "super-resonance effect" associated with internal Mie modes [13-15] and which the field structure are different from WGM. These resonances, which valid for specific values of Mie $q$-parameter, yield magnetic and electric field-intensity enhancement factors on the order up to $10^8$ [13]. Obviously, an increase in the accuracy of resonance localization (or the accuracy of the size parameter $q$) tends to increase the field enhancement limit. This provides a way to obtain field localization deep below the diffraction limit and with extreme magnetic [16] and electric field enhancement, comparable to field enhancement by subwavelength dielectric or plasmonic nanostructures [17-21], including nanohole structured sphere [22,23].

However, until now the effect of super-resonance has been studied in relation to spherical particles surrounded by vacuum ($n=1$). At the same time, considering super-resonance effect are extremely sensitive to the Mie size parameter [13], so the influence of the environment and the value of its refractive index is obviously critical.

**Super-Resolution in hot spot**

Below, we use the Mie theory, which includes a superposition of partial incident, internal and scattered fields [24-26]. The main expressions of Mie formulas allow to analytically compute both the internal and scattering fields associated to the different magnetic and electric multipolar modes supported by the particle [13]. All the results were obtained directly from analytical calculations, which reveals new physics previously unseen [13-15, 27].

In our research, as a surrounding medium, we use an air. The refractive index of air depends on many parameters, such as humidity, pressure, temperature, etc. [28-31]. In our studies, to demonstrate the effect of environmental refractive index on super resonance properties, we used the value of n=1.000241307 as example.

In the Figure 1, we illustrate the super-resonance effect for the non-absorbing mesoscale particle, immersed in vacuum, with size parameter $q$=26.94163 and refractive index $n$=1.5. These parameters correspond to a resonant mode excited inside the particle with partial wave order $l$=35 [13]. One can see two near-symmetrical hotspots were observed near the bottom and top apex points of sphere which are different from WGM structure. The enhancement factor of electric field intensity for the case of vacuum can be extremely large at these points.
However, from the comparison of two cases of surrounding medium one can immediately see that for air with $n$=1.000241307 produces around a 10 times decrease of the intensity in comparison to vacuum ($n$=1.0).

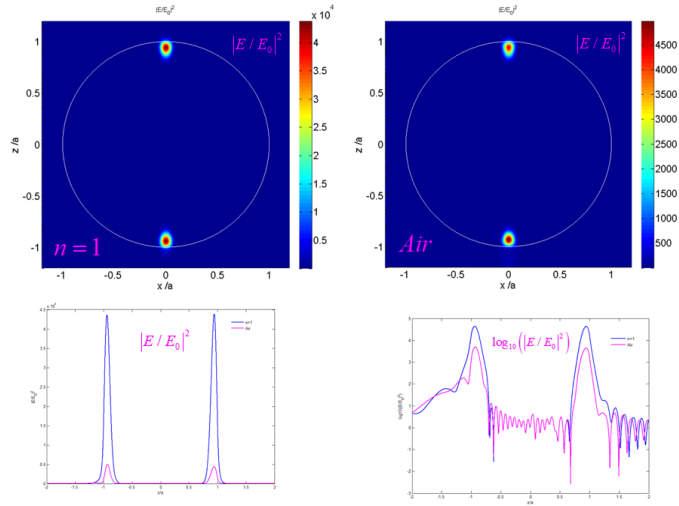

Figure 1. Hot spots generation for dielectric sphere immersed in vacuum (left) and air (right). The distribution of the electric field intensity along the hot spots (along the vertical axis) is shown below. For the case of an air environment, the graph is given on a logarithmic scale.

Figure 2 shows the corresponding distributions at the bottom hotspot, demonstrating the change in the hot spot parameters. Although changing the refractive index from vacuum to air results in a dramatic drop in electric field intensity, the focal spot size does not change much. Thus, the full width at half maximum (FWHM [32]) of the focusing spot along the major axis is FWHM=0.3742$\lambda$ for vacuum and FWHM=0.377$\lambda$ for air. For minor axis FWHM=0.2058$\lambda$ for vacuum and FWHM=0.2096$\lambda$ for air. This can be explained by the fact that the focus is on the boundary of the sphere.

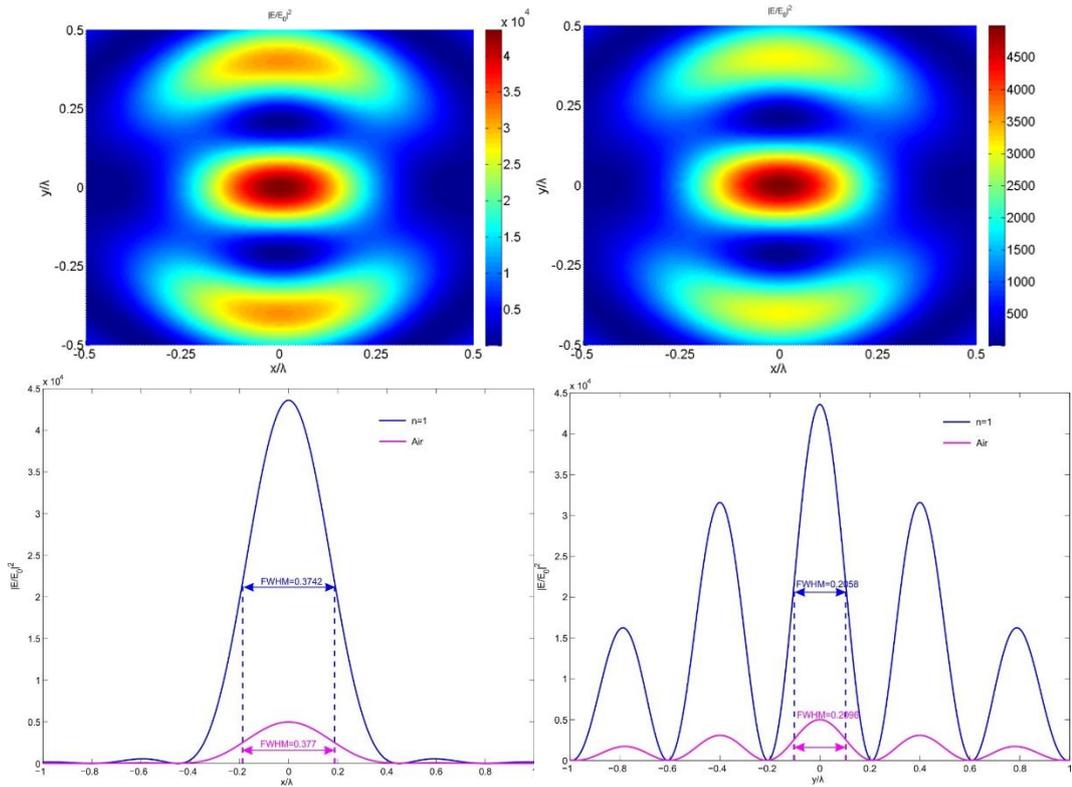

Figure 2. Electric hot spot structure for vacuum (left) and air (right) surrounding medium.

Importantly, the magnetic hotspot spot has a super resolution of 0.20-0.21λ (Fig.3), which exceeds resolution limit of WGM of 0.25λ [33-35].

One can see a small change in the refractive index of the environment (from vacuum to air) leads to a significant change in the focusing pattern. So, in the case of air, for the magnetic component of the field intensity, the hot spots degenerate, and the formation of a classical photonic jet is observed. However, a detailed analysis of the magnetic mode for other values of the Mie size parameter, due to a number of significant differences, will be considered later.

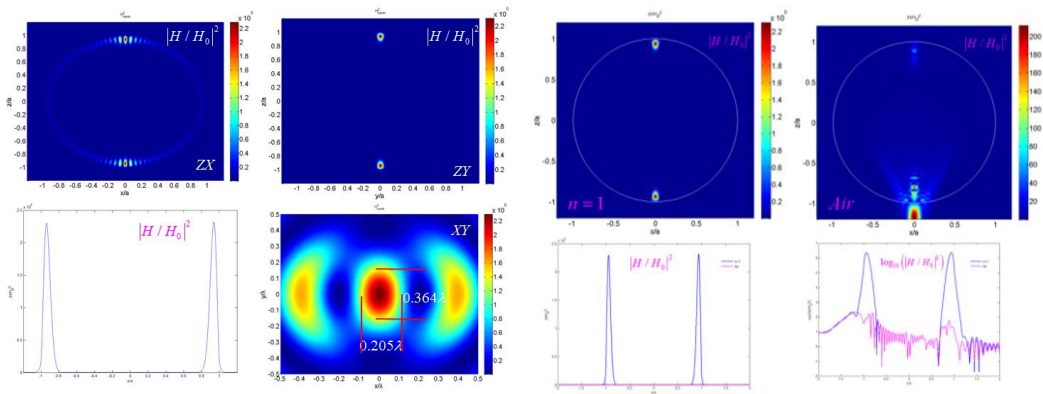

Figure 3. Magnetic hot spot generation for the sphere with q=38.6203, immersed in vacuum (left) and air (right).

**The physics of environmental influence on super-resonance effect**

Let us briefly consider the influence of the environment on the example of a sphere with Mie size parameter of *q*=26.94163.

Because super-resonance modes are excited internal partial wave modes [12,13], to analyses effect of the surrounding medium, we study the internal field amplitudes $c_l$ and $d_l$ [36, 37], which define the electric and magnetic fields inside the particle in Mie theory, similar to [12,13].

In Figure 4, we show the magnitude of the Mie coefficients $|{}^eA_l|$ and $|{}^mA_l|$ for $1 \leq l \leq 56$ and for different values of $\delta n$. One can observed from the Figure 4 that there are several visible peaks for each values of $\delta n$. There are many multipole orders for the case in this study. One can see that the resonant scattering coefficient $|{}^mA_{35}|$ is much higher in magnitude than the other coefficients and very sensitive to changes in the refractive index of the medium $\delta n$.

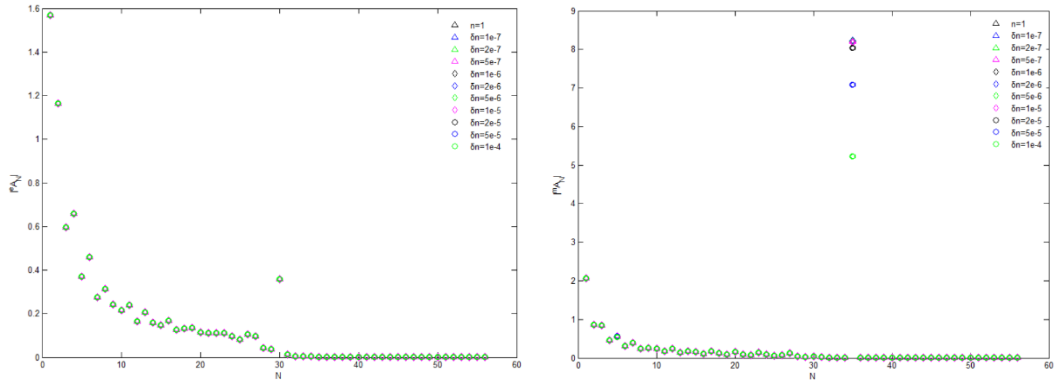

Figure 4. The first 56 Mie $|{}^eA_l|$ (left) and $|{}^mA_l|$ (right) coefficients (See equation (8) in Methods) vs different values of $\delta n$.

This very sensitive value of the scattering coefficient represents the constructive interference of the partial wave $l = 35$ (of many internal reflections) inside the mesoscale sphere.

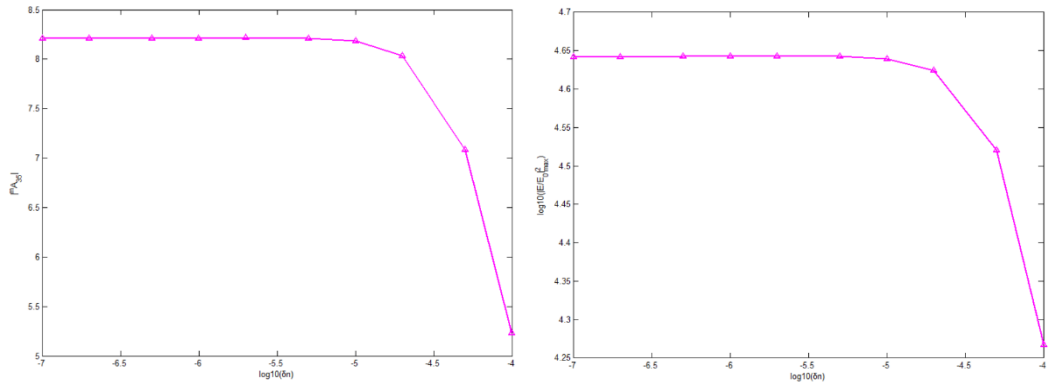

Figure 5. Dependences of the Mie resonant coefficient $|{}^mA_{35}|$ (left) and electric field intensity (right) vs different values of $\delta n$ (in log scale).

The influence of the environment on the field intensity peak can be compensated by adjusting the resonant size of the sphere. Below in Figure 6 is an example of super resonance for a sphere with an initial value of q=26.94163 surrounded by vacuum and air. Due to the small change of index, we seek the resonant q for sphere in Air around the value which is selected to make the sphere resonance in vacuum. Then, the same resonant mode is picked out to obtain the new q for sphere in Air.

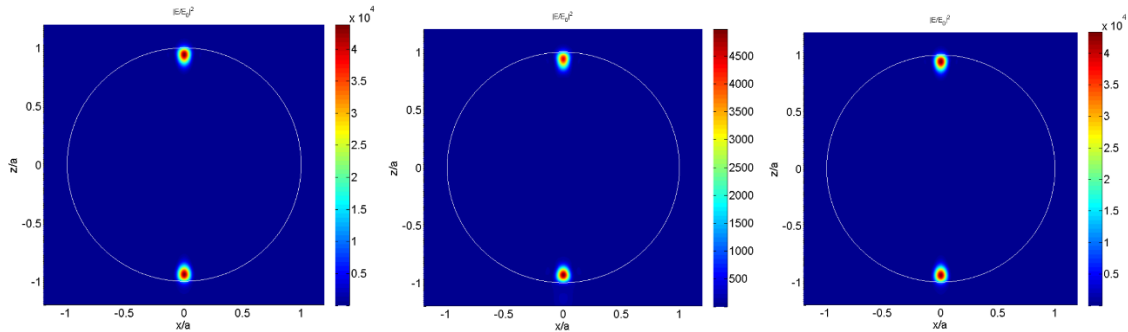

Figure 6. Super-resonant modes for sphere with q=26.94163 immersed in vacuum and air and sphere with new value of q=26.94138 immersed in air.

The corresponding field intensity distribution along the *z*-axis (vertical) is shown in Fig.7.

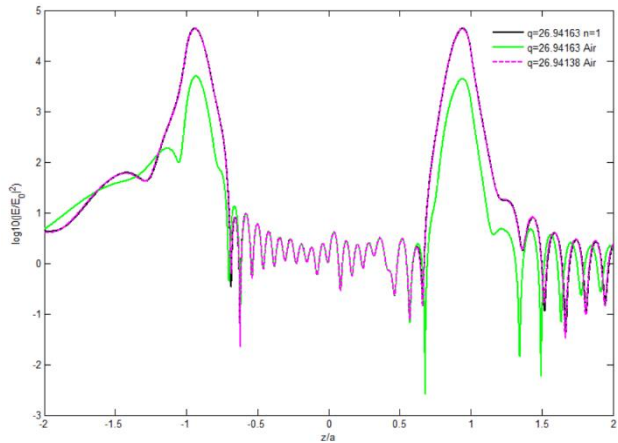

Figure 7. Electric field intensity enhancement distributions for sphere with q=26.94163 immersed in vacuum and air and sphere with new value of q=26.94138 immersed in air.

It is clearly seen that by reducing the resonant value of the sphere size parameter from q=26.94163 in vacuum to q=26.94138 in air for the example under consideration, one can completely compensate for the change in the medium index.

In order to explain this resonant size parameter shift, consider the Mie coefficients $c_l$ and $d_l$ in equations (3). The resonance is observed when the denominator in (3) tends to zero (see "Methods").

As a demonstration of the environmental effect, Figure 8 below shows the trend of the resonant size parameter *q* for a spherical particle with a refractive index of $n_s$=1.9 and mode *l*=35. The relevant main parameters are shown in Table 1.

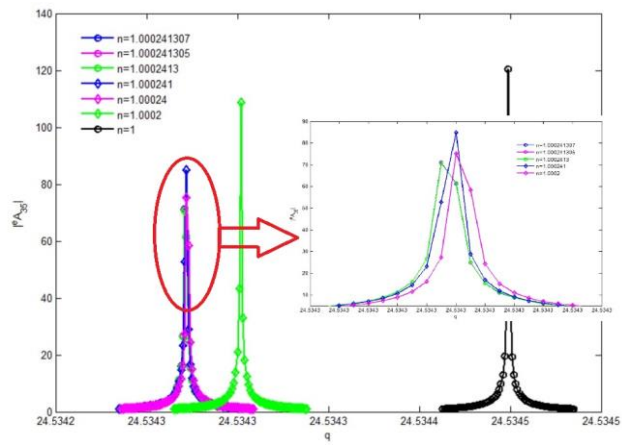

Figure 8. The displacement of the resonant value of the Mie size parameter from the change in the refractive index of the medium. Here $\left|{}^e A_{35}\right| = \left|i^{35+1} \dfrac{2*35+1}{35(35+1)} d_{35}\right|$.

Table 1.

| $n_m$ | L | q | Max ($\left|{}^e A_{35}\right|$) |
|---|---|---|---|
| 1.000241307 | 35 | 24.534271 | 71.1 |
| 1.000241305 | 35 | 24.534271 | 70.9 |
| 1.0002413 | 35 | 24.534271 | 70.6 |
| 1.000241 | 35 | 24.534272 | 84.9 |
| 1.00024 | 35 | 24.534272 | 75.2 |
| 1.0002 | 35 | 24.534302 | 108.6 |
| 1 | 35 | 24.534449 | 120.4 |

The results show that when changing the refractive index of the environment from vacuum (n=1) to air with an index of 1.0002, the resonant value of the size parameter is changed from q=24.534449 to q=24.534302. With a further increase in the refractive index of air to n=1.0002413, the resonant value of the size parameter also shifts to the "blue" region to a value of q=24.534271. With an even greater increase in the refractive index of the medium, the change in the resonant value of q is not noticeable (with the chosen accuracy in determining the "zero" of the denominator of the Mie coefficients). In this case, as can be seen from the Table 1, an increase in the refractive index of the medium leads to a decrease in the Mie coefficient about 1.7 times for above specified conditions.

The previously obtained estimates of the intensity of the magnetic and electric fields enhancement under superresonance conditions for a sphere in vacuum [13-15] can be considered as an estimate "from above".

**Conclusion**

In this work, we have investigated super-resonance modes in mesoscale dielectric spheres, immersed in vacuum and air, by using Mie theory. The results suggest that both the electric and magnetic field intensity enhancement depend essentially on a small change in the refractive index of the surrounding medium. Conditions of the super resonance effect varies as a function of refractive index contrast between particle material and surrounding medium and Mie size parameter *q*. Even a change in the refractive index in the fifth decimal place leads to a catastrophic drop in maximum intensity. In this case, with an increase in the accuracy of localization of the resonance (the value of the size parameter of the particle), one should expect an even greater influence of the environmental index. Optimization of the super resonance effect due to a more accurate localization of the dimensional parameter of the sphere without taking into account the refractive index of the medium surrounding the sphere with the appropriate accuracy is not advisable. Magnetic hotspots are more sensitive to refractive index contrast than electrical one. At the same time, the decrease in the field intensity with a changed refractive index of the medium can be compensated by changing the particle size to the blue shift.

These results are also important in understanding and explaining the super-resolution phenomena in mesoscale sphere-based techniques and microscopy [39]. Moreover, the giant field enhancement will find interesting applications in super-enhanced Raman scattering [40], refractive index sensors of surrounding medium and sphere, similar to WGM [41-43] but with more high sensitivities, extreme and non-linear optics [13, 27] and others. These effects also can be extended to the acoustic [44].

**Acknowledgements**

**Methods**
According to the Mie theory [24-26, 36], the total scattering efficiency is presented by sum of partial scattering efficiencies:

$$Q_{sca} = \sum_{\ell=1}^{\infty}\left(Q_{\ell}^{(e)} + Q_{\ell}^{(m)}\right), \quad Q_{\ell}^{(e)} = \frac{2(2\ell+1)}{q^2}|a_{\ell}|^2, \quad Q_{\ell}^{(m)} = \frac{2(2\ell+1)}{q^2}|b_{\ell}|^2. \qquad (1)$$

where each partial efficiency corresponds to the radiation of the $\ell$-th order multipole. Terms $Q_{\ell}^{(m)}$ and $Q_{\ell}^{(e)}$ describe the radiation related to the magnetic and electric polarizabilities, respectively. In the following, we will discuss transparent dielectric particles with $\operatorname{Im}\varepsilon = 0$, so $Q_{ext} = Q_{sca}$. The electric $a_{\ell}$, and magnetic $b_{\ell}$, scattering amplitudes for nonmagnetic materials with relative magnetic susceptibility $\mu = 1$, and dielectric permittivity $\varepsilon = n^2$ are given by:

$$a_{\ell} = \frac{\mathfrak{R}_{\ell}^{(a)}}{\mathfrak{I}_{\ell}^{(a)}}, \quad b_{\ell} = \frac{\mathfrak{R}_{\ell}^{(b)}}{\mathfrak{I}_{\ell}^{(b)}}, \qquad (2)$$

Electric and magnetic fields inside the particle are expressed through the internal scattering amplitudes given by

$$c_{\ell} = \frac{\mathfrak{R}_{\ell}^{(c)}}{\mathfrak{I}_{\ell}^{(c)}}, \quad d_{\ell} = \frac{\mathfrak{R}_{\ell}^{(d)}}{\mathfrak{I}_{\ell}^{(d)}}. \qquad (3)$$

where $\mathfrak{R}_{\ell}$ and $\mathfrak{I}_{\ell}$ functions are defined as follows:

$$\mathfrak{R}_{\ell}^{(a)} = y\psi_{\ell}'(x)\psi_{\ell}(y) - x\psi_{\ell}(x)\psi_{\ell}'(y), \quad \mathfrak{I}_{\ell}^{(a)} = y\chi_{\ell}'(x)\psi_{\ell}(y) - x\chi_{\ell}(x)\psi_{\ell}'(y), \qquad (4)$$

$$\mathfrak{R}_{\ell}^{(b)} = y\psi_{\ell}'(x)\psi_{\ell}(y) - x\psi_{\ell}(y)\psi_{\ell}'(x), \quad \mathfrak{I}_{\ell}^{(b)} = y\chi_{\ell}(x)\psi_{\ell}'(y) - x\psi_{\ell}(y)\chi_{\ell}'(x). \qquad (5)$$

$$\mathfrak{R}_{\ell}^{(c)} = y\psi_{\ell}'(x)\psi_{\ell}(x) - y\psi_{\ell}(x)\psi_{\ell}'(x), \quad \mathfrak{I}_{\ell}^{(c)} = y\chi_{\ell}'(x)\psi_{\ell}(y) - x\chi_{\ell}(x)\psi_{\ell}'(y), \qquad (6)$$

$$\mathfrak{R}_{\ell}^{(d)} = y\psi_{\ell}'(x)\psi_{\ell}(x) - y\psi_{\ell}(x)\psi_{\ell}'(x), \quad \mathfrak{I}_{\ell}^{(d)} = y\chi_{\ell}(x)\psi_{\ell}'(y) - x\psi_{\ell}(y)\chi_{\ell}'(x). \qquad (7)$$

Here, $\psi_{\ell}(z) = \sqrt{\frac{\pi z}{2}} J_{\ell+\frac{1}{2}}(z)$, $\chi_{\ell}(z) = \sqrt{\frac{\pi z}{2}} N_{\ell+\frac{1}{2}}(z)$, where $J_{\ell+\frac{1}{2}}(z)$ and $N_{\ell+\frac{1}{2}}(z)$ are the Bessel and Neumann functions, $x = k_m R$, $y = k_p R$, $k_{m,p} = 2\pi\sqrt{\varepsilon_{m,p}}/\lambda$ is a wavevector in medium and particle, respectively. The radius of the particle $R$ enters in this theory through the dimensionless Mie size parameter $q = 2\pi R/\lambda$, where $\lambda$ the radiation wavelength in vacuum. The prime in formulas (3), (4) indicates differentiation with respect to the argument of the function, i.e. $\psi_{\ell}'(z) \equiv d\psi_{\ell}(z)/dz$, etc. The coefficients in the equations for field's components [25] are:

$$^{l}B_{n} = i^{n+1}\frac{2n+1}{n(n+1)}a_{n}, \quad ^{m}B_{n} = i^{n+1}\frac{2n+1}{n(n+1)}b_{n}, \quad ^{l}A_{n} = i^{n+1}\frac{2n+1}{n(n+1)}c_{n}, \quad ^{m}A_{n} = i^{n+1}\frac{2n+1}{n(n+1)}d_{n}. \quad (8)$$

(Detailed equations for scattered wave both in medium and inside non-magnetic sphere maybe found, for example, in [25]).

Due to the numerators of (3) never tends to zero, the values of amplitudes $|c_\ell|^2$ and $|d_\ell|^2$ are not restricted by unity as amplitudes $|a_\ell|^2$ and $|b_\ell|^2$ in (2), but increase with refractive index contrast and values of Mie size parameter. To compare both type of resonances we introduce [13] internal scattering efficiency (not defined in classical Mie theory) $Q_{sci}^{(\ell)}$ with partial internal scattering efficiencies, in analogy to Eq. (1):

$$Q_{sci}^{(\ell)} = \sum_{\ell=1}^{\infty}\left(F_\ell^{(e)} + F_\ell^{(m)}\right), \text{ where: } F_\ell^{(e)} = \frac{2(2\ell+1)}{q^2}|c_\ell|^2, \quad F_\ell^{(m)} = \frac{2(2\ell+1)}{q^2}|d_\ell|^2. \quad (9)$$

Coefficients $c_l$ and $d_l$ can be rewritten as [45]

$$c_l = \frac{y\zeta_l(x)\psi_l'(x) - y\zeta_l'(x)\psi_l(x)}{y\zeta_l'(x)\psi_l(y) - x\zeta_l(x)\psi_l'(y)}$$

$$d_l = \frac{y\zeta_l'(x)\psi_l(x) - y\zeta_l(x)\psi_l'(x)}{y\zeta_l(x)\psi_l'(y) - x\zeta_l'(x)\psi_l(y)}, \quad (10)$$

here, $x = k_m a$ and $y = k_p a$,

$$\zeta_l(\rho) = \rho h_l^{(1)}(\rho) = \sqrt{\frac{\pi\rho}{2}}H_{l+\frac{1}{2}}^{(1)}(\rho), \quad \psi_l(\rho) = \rho j_l(\rho) = \sqrt{\frac{\pi\rho}{2}}J_{l+\frac{1}{2}}(\rho),$$

$$\zeta_l'(\rho) = \frac{\partial\zeta_l(\rho)}{\partial\rho}, \quad \psi_l'(\rho) = \frac{\partial\psi_l(\rho)}{\partial\rho},$$

Let

$$\zeta_l(\rho) = \psi_l(\rho) - i\chi_l(\rho)$$

where $\chi_l(\rho) = -\rho y_l(\rho)$ and $y_l(\rho) = \sqrt{\frac{\pi\rho}{2}}Y_{l+\frac{1}{2}}(\rho)$.

Considering

$$\psi_l(\rho)\zeta_l'(\rho) - \psi_l'(\rho)\zeta_l(\rho) = i,$$

$$\chi_l(\rho)\psi_l'(\rho) - \chi_l'(\rho)\psi_l(\rho) = 1,$$

Then

$$c_l = \frac{-iy}{F_l^{(a)} + iG_l^{(a)}},$$

$$d_l = \frac{iy}{F_l^{(b)} + iG_l^{(b)}}, \quad (11)$$

Where

$$F_l^{(a)} = y\psi_l'(x)\psi_l(y) - x\psi_l(x)\psi_l'(y),$$

$$G_l^{(a)} = -y\chi_l'(x)\psi_l(y) + x\chi_l(x)\psi_l'(y),$$
$$F_l^{(b)} = y\psi_l(x)\psi_l'(y) - x\psi_l'(x)\psi_l(y),$$
$$G_l^{(b)} = -y\chi_l(x)\psi_l'(y) + x\chi_l'(x)\psi_l(y).$$

When $|c_l| \to \infty$ we have

$$F_l^{(a)} = y\psi_l'(x)\psi_l(y) - x\psi_l(x)\psi_l'(y) = 0,$$
$$G_n^{(a)} = -y\chi_n'(x)\psi_n(y) + x\chi_n(x)\psi_n'(y) = 0,$$

$$m = \frac{y}{x} = \frac{k_p}{k_m}$$

$$m\psi_l'(x)\psi_l(mx) - \psi_l(x)\psi_l'(mx) = 0$$
$$-m\chi_l'(x)\psi_l(mx) + \chi_l(x)\psi_l'(mx) = 0$$

Let

$$S^{(a)}(x) = m\psi_l'(x)\psi_l(mx) - \psi_l(x)\psi_l'(mx),$$
$$T^{(a)}(x) = -m\chi_l'(x)\psi_l(mx) + \chi_l(x)\psi_l'(mx)$$

The resonance conditions may be found from

$$|S^{(a)}(x)| \leq \delta \text{ and } |T^{(a)}(x)| \leq \delta$$

Accordingly, when $|d_l| \to \infty$

$$F_l^{(b)} = 0, G_l^{(b)} = 0$$

Then

$$m\psi_l(x)\psi_l'(mx) - \psi_l'(x)\psi_l(mx) = 0$$
$$-m\chi_l(x)\psi_l'(mx) + \chi_l'(x)\psi_l(mx) = 0$$

Let

$$S^{(b)}(x) = m\psi_l(x)\psi_l'(mx) - \psi_l'(x)\psi_l(mx)$$
$$T^{(b)}(x) = -m\chi_l(x)\psi_l'(mx) + \chi_l'(x)\psi_l(mx)$$

And the resonance conditions may be found from

$$|S^{(b)}(x)| \leq \delta \text{ and } |T^{(b)}(x)| \leq \delta$$

Here $\delta$ can evaluate of the resonance strength.